\documentclass[10pt,journal,final,oneside,twocolumn,nofonttune]{IEEEtran}
\usepackage[compress]{cite}
\usepackage{hyperref}
\makeatletter
\renewcommand{\citepunct}{,\penalty\@m\hskip.13emplus.1emminus.1em}
\renewcommand{\citedash}{\hbox{--}\penalty\@m}
\makeatother

\usepackage{graphicx}
\usepackage{amsmath}
\usepackage{amssymb}
\usepackage{amsfonts}
\usepackage{psfrag}
\usepackage{float}
\usepackage{array}
\ifCLASSOPTIONcompsoc
\usepackage[tight,normalsize,sf,SF]{subfigure}
\else
\usepackage[tight,footnotesize]{subfigure}
\fi

\IEEEoverridecommandlockouts

\begin{document}
%
\title{
Equivalence of SLNR Precoder and RZF Precoder in Downlink MU-MIMO
Systems}

\author{
\IEEEauthorblockN{\large{Fang Yuan and Chenyang Yang}}
\vspace{0.2cm}
\IEEEauthorblockA{\\Beihang University, Beijing, China\\
Email: yuanfang@ee.buaa.edu.cn, cyyang@buaa.edu.cn\ 
}
}

\maketitle
\begin{abstract}
The signal-to-leakage-and-noise ratio (SLNR) precoder is widely used
for MU-MIMO systems in many works, and observed with improved
performance from zeroforcing (ZF) precoder. Our work proofs SLNR
precoder is completely equivalent to conventional regulated ZF (RZF)
precoder, which has significant gain over ZF precoder at low SNRs.
Therefore, with our conclusion, the existing performance analysis
about RZF precoder can be readily applicable to SLNR precoder.
\end{abstract}

\section{Introduction}
Downlink Multi-user Multiple-Input-and-Multiple-Output (MU-MIMO)
techniques have attracted much attention in the last decades
\cite{Haustein2002,Jindal2006,Peel2005}. It is promising to increase
the system throughput in the future industrial standards, such as
LTE-A and WiMax.

To support multiple user in downlink, it is important to suppress
interference between users. As is known, the sum capacity of
downlink MU-MIMO can be achieved by using Dirty-Paper-Coding (DPC)
techniques \cite{Costa1983}, which is however with huge complexity.
To reduce the complexity, various linear precoders are proposed
\cite{Haustein2002,Peel2005}.

Among all, the Zeroforcing (ZF) based precoders by using channel
inversions are widely used in MU-MIMO systems because of low
complexity \cite{Haustein2002,Jindal2006}. These schemes impose a
restriction on the system that the interference between users are
cancellated. To avoid the poor performance of pure ZF precoder,
regulated ZF precoder was proposed in \cite{Peel2005}.

Different from ZF based precoders, a leakage-based precoder proposed
in \cite{Sadek2007}. The precoders are designed based on the concept
of signal ``leakage'', which refers to the interference caused by
the signal intended for a desired user to the remaining users. Such
precoder is to maximize the signal-to-leakage-and-noise ratio (SLNR)
for all users. It was observed in \cite{Sadek2007} SLNR precoder has
significant improvement from ZF precoder. A comprehensive comparison
between ZF, MRT and SLNR precoders can be found in
\cite{Bjornson2010} the comparison between SLNR and ZF that I have
madeThe SLNR precoder is with decoupled nature analytical closed
form and thus widely adopted in many optimization applications
\cite{Sharma2006,Cheng2010}.

In this work, we proof that the SLNR precoder, although inspired
from different idea, is equivalent to the conventional RZF precoder.
The equivalence not only explains why SLNR precoder is observed to
be better than ZF precoder, but also makes the existing analysis of
RZF precoder readily available to SLNR precoder.

\section{System Model}
\subsection{Downlink channel}
We consider a downlink MU-MIMO system where a BS with $N_t$ antennas
serves $K$ single-antenna users. The received signal at the $k$th
user can be expressed as
\begin{equation}
       y_k =\mathbf{h}^H_k\mathbf{w}_ks_k +\sum_{j=1, j\neq k}^{K}\mathbf{h}^H_k\mathbf{w}_js_j + n_k,\label{bf}
\end{equation}
where $\mathbf{h}_k\in \mathbb{C}^{N_t\times1}$ is the channel
vector of the $k$th user, $\mathbf{w}_k \in \mathbb{C}^{N_t\times1}$
is a unit-norm precoder, $s_k$ is the data symbol with unit variance
destined to user $k$, and $n_k$ is the additive white Gaussian noise
with zero mean and variance $\sigma^2$. With loss of generalities,
we assume $\sigma^2=1$. $()^H$ denotes the conjugate-transpose
operation.

\subsection{Existing ZF, RZF, \& SLNR precoder}
In downlink MU-MIMO systems, Zeroforcing (ZF) precoder is chosen
such that the interference between users is nulled, i.e.,
\begin{equation}
\mathbf{h}^H_k\mathbf{w}_j = 0,
\end{equation}
when $k\neq j$.

A pseu-inverse based ZF precoder $\mathbf{w}_{\text{ZF},k}$
\cite{Haustein2002} is as
\begin{equation}
\mathbf{w}_{\text{ZF},k} \propto
(\mathbf{H}\mathbf{H}^H)^{-1}\mathbf{h}_k.
\end{equation}
where $\propto$ means linear proportionality, and $\mathbf{H}=
[\mathbf{h}_1, \cdots, \mathbf{h}_K]$.

A regulated ZF precoder (RZF) is proposed in \cite{Peel2005} to
improve pure ZF precoder as
\begin{equation}
\mathbf{w}_{\text{RZF},k} \propto
(\alpha\mathbf{I}+\mathbf{H}\mathbf{H}^H)^{-1}\mathbf{h}_k,\label{MMSE}
\end{equation}
where $\alpha$ is the regulation parameter. It is worthy to note
that optimal $\alpha$ varies with system configurations. However, a
most frequently choice is $\alpha=\sigma^2$, as recommended in
\cite{Peel2005}, which is optimal when $K$ is large and works well
even when $K$ is small.

Different from previous precoders, the signal-to-leakage-noise ratio
(SLNR) precoder adopts a leakage-base solution. The SLNR of single
stream is defined as
\begin{align}
\text{SLNR}_k &= \frac{|\mathbf{h}^H_k\mathbf{w}_k|^2}{1+\sum_{j\neq
k}|\mathbf{h}^H_j\mathbf{w}_k|^2}\nonumber\\
&= \frac{|\mathbf{h}^H_k\mathbf{w}_k|^2}{1+
\|\mathbf{H}^H_{-k}\mathbf{w}_k\|^2}\label{SLNRF}
\end{align}
where $\mathbf{H}_{-k}= [\mathbf{h}_1,
\cdots,\mathbf{h}_{k-1},\mathbf{h}_{k+1}, \cdots, \mathbf{h}_K]$.
The denominator in the above equation defines the sum of the noise
power and the total interference power leaked from one user to the
other users.

The SLNR precoder to maximize the SLNR in \eqref{SLNRF} is as
\cite{Sadek2007}
\begin{equation}
\mathbf{w}_{\text{SLNR},k} = \text{max. eigenvector} \left(
(\mathbf{I}+\mathbf{H}_{-k}\mathbf{H}^H_{-k})^{-1}\mathbf{h}_k\mathbf{h}^H_k
\right)\label{SLNR}
\end{equation}

In the case of multiple receiver antennas at the user, extensions of
the SLNR precoder from single stream to multi stream can refer to
\cite{Cheng2010}. Our work focus on single stream case and multi
stream case can be investigated in further work. This approach is
also supported by the prevalent LTE-standards, where each stream is
treated as an individual users irrespective of multiple receiver
antennas.

\section{Equivalence of SLNR and RZF precoder}

We will show that the SLNR precoder is equivalent to RZF precoder.
To do this, we first introduce the following lemma.

\newtheorem{lemma}{Lemma}
\begin{lemma}\label{colly1}
The two vectors satisfy
$(\mathbf{I}+\mathbf{H}_{-k}\mathbf{H}^H_{-k})^{-1}\mathbf{h}_k
\propto (\mathbf{I}+\mathbf{H}\mathbf{H}^H)^{-1}\mathbf{h}_k$.
\end{lemma}
\vspace{2mm}

{Proof}: From matrix inverse identity \cite{VT02}
\begin{equation}
(\mathbf{A}+\mathbf{x}\mathbf{x}^H)^{-1} = \mathbf{A}^{-1}-
\frac{\mathbf{A}^{-1}\mathbf{x}\mathbf{x}^H\mathbf{A}^{-1}}{1+\mathbf{x}^H\mathbf{A}^{-1}\mathbf{x}},
\end{equation}
we have
\begin{align}
(\mathbf{A}+\mathbf{x}\mathbf{x}^H)^{-1}\mathbf{x} &=
\mathbf{A}^{-1}\mathbf{x}-
\frac{\mathbf{A}^{-1}\mathbf{x}\mathbf{x}^H\mathbf{A}^{-1}}{1+\mathbf{x}^H\mathbf{A}^{-1}\mathbf{x}}\mathbf{x}\nonumber\\
&= \mathbf{A}^{-1}\mathbf{x}-
\frac{\mathbf{x}^H\mathbf{A}^{-1}\mathbf{x}}{1+\mathbf{x}^H\mathbf{A}^{-1}\mathbf{x}}\mathbf{A}^{-1}\mathbf{x}\nonumber\\
&=
\frac{1}{1+\mathbf{x}^H\mathbf{A}^{-1}\mathbf{x}}\mathbf{A}^{-1}\mathbf{x},
\end{align}
which means
\begin{align}
(\mathbf{A}+\mathbf{x}\mathbf{x}^H)^{-1}\mathbf{x} \propto
\mathbf{A}^{-1}\mathbf{x}.
\end{align}

By setting $\mathbf{A}=\mathbf{I}+\mathbf{H}_{-k}\mathbf{H}^H_{-k}$
and $\mathbf{x}=\mathbf{h}_k$,  we immediately obtain Lemma
1.$\blacksquare$

\vspace{2mm}
\newtheorem{therom}{Therom}
\begin{therom}\label{th1}
The SLNR precoder obtained in \eqref{SLNR} is equivalent to the RZF
precoder obtained in \eqref{MMSE}.
\end{therom}
\vspace{2mm}

{Proof}: Define $\tilde{\mathbf{w}}_k =\frac{1}{\sqrt{\gamma}}
(\mathbf{I}+\mathbf{H}_{-k}\mathbf{H}^H_{-k})^{-1}\mathbf{h}_k$,
where $\gamma=
\|(\mathbf{I}+\mathbf{H}_{-k}\mathbf{H}^H_{-k})^{-1}\mathbf{h}_k\|^2$.
We observe that
\begin{align}
&\hspace{1em}(\mathbf{I}+\mathbf{H}_{-k}\mathbf{H}^H_{-k})^{-1}\mathbf{h}_k\mathbf{h}^H_k\tilde{\mathbf{w}}_k\nonumber\\
&=(\mathbf{I}+\mathbf{H}_{-k}\mathbf{H}^H_{-k})^{-1}\mathbf{h}_k\mathbf{h}^H_k\frac{1}{\sqrt{\gamma}}
(\mathbf{I}+\mathbf{H}_{-k}\mathbf{H}^H_{-k})^{-1}\mathbf{h}_k\nonumber\\
&=\mathbf{h}^H_k\frac{1}{\sqrt{\gamma}}
(\mathbf{I}+\mathbf{H}_{-k}\mathbf{H}^H_{-k})^{-1}\mathbf{h}_k
(\mathbf{I}+\mathbf{H}_{-k}\mathbf{H}^H_{-k})^{-1}\mathbf{h}_k\nonumber\\
&\triangleq  \lambda \tilde{\mathbf{w}}_k
\end{align}
where $\lambda=\mathbf{h}^H_k
(\mathbf{I}+\mathbf{H}_{-k}\mathbf{H}^H_{-k})^{-1}\mathbf{h}_k>0$.
It indicates that $\tilde{\mathbf{w}}_k$ is an eigenvector of
$(\mathbf{I}+\mathbf{H}_{-k}\mathbf{H}^H_{-k})^{-1}\mathbf{h}_k\mathbf{h}^H_k$.

Moreover, due to the matrix product rank inequality
$\text{rank}(\mathbf{AB})\leq
\min\{\text{rank}(\mathbf{A}),\text{rank}(\mathbf{B})\}$, we find
that
$\text{rank}\left((\mathbf{I}+\mathbf{H}_{-k}\mathbf{H}^H_{-k})^{-1}\mathbf{h}_k\mathbf{h}^H_k\right)\leq
\text{rank}\left(\mathbf{h}_k\mathbf{h}^H_k\right)=1$. Because
$(\mathbf{I}+\mathbf{H}_{-k}\mathbf{H}^H_{-k})^{-1}\mathbf{h}_k\mathbf{h}^H_k$
is a nonzero matrix, we have
\begin{align}
\text{rank}\left((\mathbf{I}+\mathbf{H}_{-k}\mathbf{H}^H_{-k})^{-1}\mathbf{h}_k\mathbf{h}^H_k\right)=1.
\end{align}

Since there is only one eigenvector, we know that
$\tilde{\mathbf{w}}_k=\text{max. eigenvector} \left(
(\mathbf{I}+\mathbf{H}_{-k}\mathbf{H}^H_{-k})^{-1}\mathbf{h}_k\mathbf{h}^H_k
\right)$.

From Lemma 1,  $\mathbf{w}_{\text{SLNR},k}=\tilde{\mathbf{w}}_k
\propto \mathbf{w}_{\text{RZF},k}$. Due to unit norm constraint, we
finally conclude that $\mathbf{w}_{\text{SLNR},k} =
\mathbf{w}_{\text{RZF},k}$.$\blacksquare$ \vspace{2mm}

To this end, we show that SLNR precoder, though inspired from
different precoding strategy, is equivalent to conventional RZF
precoder.

\section{Discussions}
The reason that SLNR precoder outperform ZF precoder can be
consequently illustrated. It is well known that RZF precoder can
significantly increase downlink MU-MIMO systems from pure ZF
precoder at low SNRs. In fact, this is exactly analogous to the
difference between ZF equalization and minimum mean-square error
(MMSE) equalization: while zero-forcing results in complete
cancelation of user interference, an MMSE equalizer allows a
measured amount of interference such that the output SNR is
maximized.

By using the equivalence of SLNR and RZF precoder, existing works
about RZF precoder can be readily applied to SLNR precoder. For
example, to get performance analysis of SLNR precoder, details can
be found in \cite{Peel2005}, where RZF precoder is analyzed.

It is shown in \cite{Sadek2007} the SLNR precoder does not require
that transmit antenna number should be no less than the number of
streams supported in the downlink. However, it can be expected that
even SLNR precoder working in this configuration will result in
severe interference between users, which is obivous for RZF
precoders.

\section{Simulation Results}
In this section, we assume that the channel is subjected to i.i.d.
flat Rayleigh fading. Moreover, we assume homogeneous users in the
downlink where users are with equal SNRs and the transmit power is
equally allocated to each users.

In the first simulation, we investigate the sum rate of MU-MIMO
systems versus different SNRs with different precoders: ZF, SLNR and
RZF. The rate is calculated using the Shannon's formula by assume
perfect link adaptation. Two configurations are considered. One is
$N_t=K=4$, and the other is $N_t=K=2$. As shown in the figure, in
spite of different duplexing gain, we find that in both two cases
the sum rate of MU-MIMO system using SLNR precoder is the same with
that of using RZF precoder. SLNR/RZF precoder outperforms that of
using ZF at low SNRs and converges to ZF precoder at high SNRs. The
simulation results support our analysis.

\begin{figure}[htb]
  \centering
  \centerline{\includegraphics[width=8.5cm]{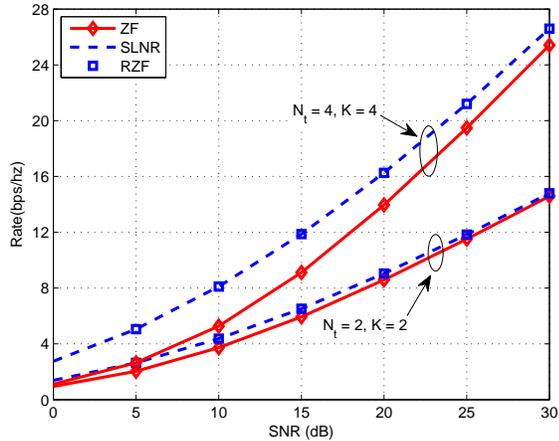}}
\caption{Sum rate of MU-MIMO systems} \label{aqam}
\end{figure}

In the second simulation, we investigate the bit error rate (BER) of
MU-MIMO systems using 4 QAM with different precoders: ZF, SLNR and
RZF. Two configurations are considered. We set $K=4$ and in the
first configuration $N_t=4$, and in the second is $N_t=6$. As shown
in the figure, the diversity order of BER in the second case is
higher than that in the first one. In both cases, we find that the
BER of MU-MIMO system using SLNR precoder is the same with that of
using RZF precoder, outperforming ZF precoder. Similar results can
be observed from other configurations where our analysis is still
valid.

\begin{figure}[htb]
  \centering
  \centerline{\includegraphics[width=8.5cm]{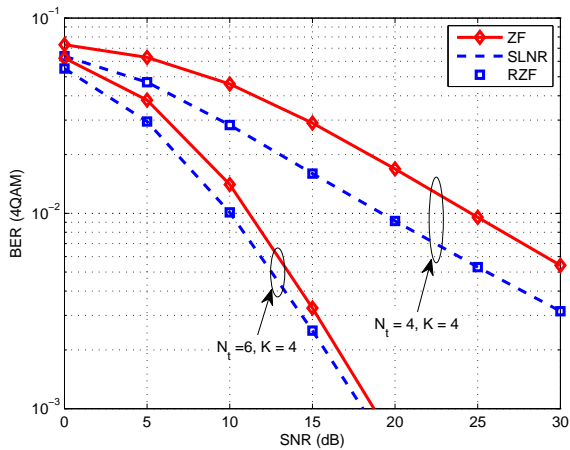}}
\caption{BER of MU-MIMO systems with 4 QAM} \label{distortionfigure}
\end{figure}

\section*{Conclusions}
In this work, we have proofed that the SLNR precoder is equivalent
to conventional regulated ZF precoder, which illustrates why SLNR
precoder outperforms ZF precoders in many applications. Our work
makes the previous analysis of RZF precoders available for SLNR
precoders.

\bibliographystyle{IEEEtran}
\bibliography{IEEEabrv,strings,Reference}
\end{document}